%% file: pptth_radcor.tex
\documentclass{PoS}

\usepackage[T1]{fontenc}
\usepackage{booktabs}
\usepackage[dvips,table]{xcolor} 
\usepackage{xspace}
\usepackage{dcolumn}
\usepackage{caption}
\usepackage
[subrefformat=parens,position=top,skip=-15pt,margin=15pt,justification=justified,singlelinecheck=false]
{subcaption}
\usepackage{comment}
\usepackage{amsmath}

\include{macros}

\title{Precise predictions for Higgs-boson production in association with top quarks}

\ShortTitle{Precise predictions for Higgs production in association with top quarks}

\author{\speaker{Ansgar Denner}\\ 
        Universit\"at W\"urzburg, Institut f\"ur Theoretische Physik und Astrophysik,
        D-97074 W\"urzburg, Germany\\
        E-mail: \email{denner@physik.uni-wuerzburg.de}}

\author{Robert Feger\\
        Universit\"at W\"urzburg, Institut f\"ur Theoretische Physik und Astrophysik,
        D-97074 W\"urzburg, Germany\\
        E-mail: \email{robert.feger@gmail.com}}

\author{Andreas Scharf\\
        Universit\"at W\"urzburg, Institut f\"ur Theoretische Physik und Astrophysik,
        D-97074 W\"urzburg, Germany\\
        E-mail: \email{ascharf@physik.uni-wuerzburg.de}}

      \abstract{ We analyse the production of a Higgs boson in
        association with a top--antitop-quark pair in the Standard
        Model at the LHC.  Considering the final state consisting of
        four b jets, two jets, one identified charged lepton and
        missing energy, we examine the irreducible background for the
        production rate and several kinematical distributions. While
        ttH production and decay is roughly a fourth of the full
        process for the final state specified above, ttbb production
        constitutes the main contribution with about $92\%$.
        Surprisingly, interference effects result in a reduction of
        the cross-section by five per cent.  Furthermore, we consider
        NLO QCD corrections for the production of a Higgs boson, two
        charged leptons, two neutrinos, and two b jets.  We discuss
        the size of the corrections and the scale dependence for the
        integrated cross section and different distributions.  For the
        integrated cross section we find a $K$ factor of 1.17 and a
        reduction of the scale dependence from $30\%$ at leading order
        to $5\%$ at next-to-leading order. }

\FullConference{12th International Symposium on Radiative Corrections (Radcor 2015) and LoopFest XIV (Radiative Corrections for the LHC and Future Colliders)\\
                15-19 June, 2015\\
                UCLA Department of Physics \& Astronomy Los Angeles, USA}

\begin{document}

\section{Introduction}
\label{sec:introduction}

After the discovery of a Higgs boson with a mass around 125\GeV by the
CMS and ATLAS collaborations \cite{Aad:2012tfa,Chatrchyan:2012ufa},
its properties need to be precisely investigated using all accessible
Higgs production and decay modes.  The production of a
Higgs boson in association with a top-quark pair is of particular
interest as it allows to directly access the top-quark Yukawa
coupling.  In the ongoing run 2 of the LHC the determination of the
$\ttbarh$ signal and the potential measurement of the top-quark Yukawa
coupling will be pursued and corresponding theoretical predictions are
required.

Next-to-leading-order (NLO) QCD corrections for the production of a
top--antitop pair in association with a Higgs boson have been calculated in
\citeres{Beenakker:2001rj,Beenakker:2002nc,Reina:2001sf,Dawson:2002tg,
  Dawson:2003zu} and matched to parton
showers \cite{Frederix:2011zi,Garzelli:2011vp,Hartanto:2015uka}. 
Recently electroweak corrections to $\Pt\bar\Pt\PH$ production have
been computed \cite{Frixione:2014qaa,Yu:2014cka,Frixione:2015zaa}.
NLO QCD corrections for the important background processes
$\Pt\bar\Pt\Pb\bar\Pb$ and $\Pt\bar\Pt\Pj\Pj$ production have been
worked out in
\citeres{Bredenstein:2008zb,Bredenstein:2009aj,Bevilacqua:2009zn,Bredenstein:2010rs}
and \citeres{Bevilacqua:2010ve,Bevilacqua:2011aa,Bevilacqua:2014qfa},
respectively, and matched to parton showers in
\citeres{Kardos:2013vxa,Cascioli:2013era,Garzelli:2014aba} and
\citere{Hoeche:2014qda}.  
In all these calculations the top quarks
and the Higgs boson have been treated as stable particles.

In these proceedings we summarise a leading-order study of Higgs-boson
production in association with a top-quark pair ($\ttbarh$) including
the subsequent semileptonic decay of the top-quark pair and the decay
of the Higgs boson into a bottom--antibottom-quark pair,
\ttbarhProcess\ \cite{Denner:2014wka}.  We consider this process in
three different scenarios, 1) the full process with all Standard Model
(SM) Feynman diagrams for the 8-particle final state, 
2) $\ttbarbbbar$ production, where only diagrams with resonant
top--antitop-quark pairs are taken into account, and 3) $\ttbarh$
production, where in addition a resonant Higgs boson is required. 
Comparing the predictions in the three scenarios allows us to examine
the size of the irreducible background for Higgs production in
association with a top--antitop-quark pair.  We have in particular
studied different methods of assigning a b-jet pair to the Higgs boson
and compared their performance in reconstructing the Higgs signal.
Furthermore, we have investigated the size of interference effects
between contributions to the matrix elements of different order in the
strong and electroweak coupling constants.

We also report on a calculation of the NLO QCD corrections to the
hadronic production of a positron, a muon, missing energy, two
\Pb~jets and a SM Higgs boson, $\fullProcessH$ \cite{Denner:2015yca},
which includes the resonant production of a top--antitop-quark pair in
association with a Higgs boson with a subsequent leptonic decay of the
top and the antitop quark.  Our calculation includes all NLO QCD
effects in $\Pt\bar\Pt\PH$ production and top decays and also takes
into account all off-shell, non-resonant and interference effects of
the top quarks.

\section{Calculational framework}

The full LO process $\fullProcess$ involves partonic channels with up
to 78,000 diagrams. All matrix elements are calculated with {\sc
  Recola}~\cite{Actis:2012qn} which provides a fast and numerically
stable computation. RECOLA uses recursive methods and allows to
specify intermediate particles for a given process. The phase-space
integration is performed with an in-house multi-channel Monte-Carlo
program, using phase-space mappings similar to
\citere{Dittmaier:2002ap}. 

We use the complex-mass scheme
\cite{Denner:1999gp,Denner:2005fg,Denner:2006ic} for the consistent
description of all resonances that are not treated in the pole
approximation.

We investigate the cross section and differential distributions for
the LHC operating at $13\TeV$.  We employ LHAPDF 6.05 with CT10 parton
distributions and neglect contributions from the suppressed
bottom-quark parton density and flavour mixing. The electromagnetic
coupling $\alpha$ is derived from the Fermi constant in the $G_\mu$
scheme. The width of the top quark \Gt is calculated at LO and NLO
QCD including effects of off-shell W bosons according to
\citere{Jezabek:1988iv}.

We use a standard set of acceptance cuts:
\begin{equation}\label{eqn:cuts}
        \begin{aligned}
                \text{non-b jets:}                        && p_{\text{T},\Pj}         &>  25\GeV,  & |y_\Pj| &< 2.5, \hspace{15ex}\\
                \text{b jets:}                      && p_{\text{T},\Pb}         &>  25\GeV,  & |y_\Pb| &< 2.5,              \\
                \text{charged lepton:}              && p_{\text{T},\Pl^+  }         &>  20\GeV,  & |y_{\Pl^+}|   &< 2.5,              \\
                \text{missing transverse momentum:} && p_{\text{T},\text{miss}} &>  20\GeV,                                        \\
                \text{jet--jet distance:}           && \Delta R_{\Pj\Pj}        &> 0.4,                                            \\
                \text{b-jet--b-jet distance:}       && \Delta R_{\Pb\Pb}        &> 0.4,                                            \\
                \text{jet--b-jet distance:}         && \Delta R_{\Pj\Pb}        &> 0.4.                               
        \end{aligned}
\end{equation}

\section{Irreducible background and interference effects in $\mathbf{\fullProcess}$}

We consider three scenarios to calculate the process $\fullProcess$:
\begin{itemize}
\item In the first scenario, the \textit{full process}, we include all SM
  contributions to the process $\fullProcess$. 
  Matrix elements involving
  external gluons receive contributions of $\order{\alphas\alpha^3}$,
  $\order{\alphas^2\alpha^2}$ and $\order{\alphas^3\alpha}$, whereas
  amplitudes without external gluons receive an additional
  $\order{\alpha^4}$ term of pure electroweak origin. 
\item In the second scenario we only take those diagrams into account
  that contain an intermediate top--antitop-quark pair. The resulting
  amplitude, labelled \textit{$\ttbarbbbar$ production} in the
  following, corresponds to the production of a bottom--antibottom
  pair and an intermediate top--antitop pair followed by its
  semileptonic decay, i.e. $\ttbarProcess$. 
  Note that we use the pole approximation
  \cite{Kleiss:1988xr,Stuart:1991xk,Aeppli:1993rs} for the top quarks only,
  hence we take into account all off-shell effects of the remaining
  unstable particles. 
  As a consequence of the
  required top--antitop-quark pair the amplitudes receive no
  contribution of $\order{\alphas^3\alpha}$.
\item Finally, we consider the signal process $\ttbarhProcess$ and
  label it \textit{$\ttbarh$ production}. In addition to the
  intermediate top--antitop-quark pair we require an intermediate
  Higgs boson decaying into a bottom--antibottom-quark pair and use
  the pole approximation for the top-quark pair and the Higgs boson.
  The requirement of the Higgs boson eliminates
  contributions of $\order{\alphas^2\alpha^2}$ from the amplitude.
\end{itemize}

%
In this analysis we take the bottom quarks massive and use the fixed
renormalization and factorization scale according to
\citere{Beenakker:2002nc},
\begin{equation}\label{eqn:FixedScale}
        \mu_\text{fix} = \mu_\text{R} = \mu_\text{F} = \frac{1}{2}\left(2m_\Pt + m_\PH\right) = 236\GeV.
\end{equation}

In
\reftas{table:results_ttxh_onshell_projected}--\ref{table:results_full_matrixelement_summary}
we present individual contributions to the integrated cross section
for the three scenarios.  While the first column specifies the
partonic initial states ($q=\Pu,\Pd,\Pc,\Ps$), the following columns
contain the contributions resulting from the square of matrix elements
of specific orders in the strong and electroweak coupling.  The column
labelled ``Sum'' represents the sum of the preceding columns, whereas
the last column labelled ``Total'' provides the integrated cross
section including in addition all interference effects between
different orders in the couplings.

 In
\refta{table:results_ttxh_onshell_projected} we show the cross
section for \ttbarh production and the corresponding contributions
resulting from quark--antiquark annihilation and gluon fusion.
\begin{table}
        \centering
        \renewcommand\arraystretch{1.2}
        \rowcolors{1}{}{tablerowcolor}
        \begin{tabular}{ld{1.9}d{2.7}d{1.7}}
                \toprule\rowcolor{tableheadcolor}       
                \textbf{pp}            
                & \multicolumn{3}{>{\columncolor{tableheadcolor}}l}{\textbf{Cross section [fb]}}\\\rowcolor{tableheadcolor}
                & \multicolumn{1}{>{\columncolor{tableheadcolor}}c}{\boldmath$\order{(\alpha^4)^2}$}%
                & \multicolumn{1}{>{\columncolor{tableheadcolor}}c}{\boldmath$\order{(\alphas\alpha^3)^2}$}%
                & \multicolumn{1}{>{\columncolor{tableheadcolor}}c}{\bf Total}\\
                \midrule
                $q \bar{q}$ & 0.014887(2)                                            & 2.1467(2)    &  2.1621(2) \\
                $\Pg \Pg$   & \multicolumn{1}{>{\columncolor{tablerowcolor}}l}{--} & 5.230(1)     &  5.2298(9) \\\midrule\rowcolor{tableheadcolor}  
                $\sum$      & 0.014887(2)                                            & 7.377(1)     &  7.3920(9) \\
                \bottomrule
        \end{tabular}
        \caption{\label{table:results_ttxh_onshell_projected} 
                Composition of the cross section in fb for \ttbarh production at the 
                LHC at $13\TeV$.  
        }
\end{table}
About $70\,\%$ of the events originate
from the gluon-fusion process. While the bulk of the contributions
results from matrix elements of order $\order{\alphas\alpha^3}$,
quark--antiquark annihilation receives an additional tiny
contribution from pure electroweak interactions. Note that there are
no interferences between diagrams of $\order{\alpha^4}$ and
$\order{\alphas\alpha^3}$ in this scenario.

The composition of the cross section for $\ttbarbbbar$ production is
shown in \refta{table:results_ttxbbx_onshell_projected}.
\begin{table}
        \centering
        \renewcommand\arraystretch{1.2}
        \rowcolors{1}{}{tablerowcolor}
        \begin{tabular}{ld{1.10}d{2.7}d{2.7}d{1.7}d{1.7}}
                \toprule\rowcolor{tableheadcolor}       
                \textbf{pp}            
                & \multicolumn{5}{>{\columncolor{tableheadcolor}}l}{\textbf{Cross section [fb]}}\\\rowcolor{tableheadcolor}
                & \multicolumn{1}{>{\columncolor{tableheadcolor}}c}{\boldmath$\order{(\alpha^4)^2}$}%
                & \multicolumn{1}{>{\columncolor{tableheadcolor}}c}{\boldmath$\order{(\alphas\alpha^3)^2}$}%
                & \multicolumn{1}{>{\columncolor{tableheadcolor}}c}{\boldmath$\order{(\alphas^2\alpha^2)^2}$}%
                & \multicolumn{1}{>{\columncolor{tableheadcolor}}c}{\bf Sum}%
                & \multicolumn{1}{>{\columncolor{tableheadcolor}}c}{\bf Total}\\
                \midrule
                $q \bar{q}$ &  0.018134(6)                                          &  2.4932(9) & 0.9199(2) &   3.4312(9) &   3.4366(6) \\
                $\Pg \Pg$   &  \multicolumn{1}{>{\columncolor{tablerowcolor}}l}{--} &  7.818(4)  & 16.650(9)  &  24.47(1)   &  23.010(7)  \\\midrule\rowcolor{tableheadcolor} 
                $\sum$      &  0.018134(6)                                          & 10.311(4)  & 17.570(9)  &  27.90(1)   &  26.446(7)  \\
                \bottomrule
        \end{tabular}
        \caption{\label{table:results_ttxbbx_onshell_projected}  
                Composition of the cross section in fb for
                $\ttbarbbbar$ production at the LHC at $13\TeV$. 
        }
\end{table}
We find a significant enhancement of the production rate compared to
\ttbarh production, and thus the irreducible background $\sigma^{\rm
  Irred.}_{\ttbarbbbar} = \sigma^{\rm Total}_{\ttbarbbbar} -
\sigma^{\rm Total}_{\ttbarh} = 19.06\fb$ exceeds the \ttbarh signal by
a factor of 2.6. The major contribution to the irreducible background
is due to QCD production of $\order{(\alphas^2\alpha^2)^2}$.  The
additional contributions of $\order{(\alphas\alpha^3)^2}$ in the
$\ttbarbbbar$~scenario result from Feynman diagrams involving
electroweak interactions with $\PZ$~bosons, $\PW$~bosons and photons,
where in particular $\Pt\bar{\Pt}\PZ$ production contributes
$1.01\fb$.
The difference between the fifth (Sum) and sixth (Total) column in
\refta{table:results_ttxbbx_onshell_projected} is due to interference
contributions between matrix elements of different orders in the
coupling constants. These cause a reduction of the cross sections by
about $5\,\%$.  The dominant effect is due to interferences of
diagrams of $\order{\alphas\alpha^3}$ where a W~boson is exchanged in the
$t$-channel with diagrams of $\order{\alphas^2\alpha^2}$ that yield
the dominant irreducible background. These kinds of interferences are
absent in the $\qqb$ channel.  On the other hand, we found the
interference of the $\Pt\bar\Pt\PH$ signal process with the dominant
irreducible background of order $\order{\alphas^2\alpha^2}$ to be
below one per cent.

\begin{table}
        \centering
        \renewcommand\arraystretch{1.2}
        \setlength{\tabcolsep}{5pt}
        \rowcolors{1}{}{tablerowcolor}
        \begin{tabular}{ld{1.9}d{2.7}d{1.7}d{1.8}d{1.7}d{2.6}d{2.6}}
                \toprule\rowcolor{tableheadcolor}       
                \textbf{pp}                 
                & \multicolumn{6}{>{\columncolor{tableheadcolor}}l}{\textbf{Cross section [fb]}}\\\rowcolor{tableheadcolor}
                & \multicolumn{1}{>{\columncolor{tableheadcolor}}c}{\boldmath$\order{(\alpha^4)^2}$}%
                & \multicolumn{1}{>{\columncolor{tableheadcolor}}c}{\boldmath$\order{(\alphas\alpha^3)^2}$}%
                & \multicolumn{1}{>{\columncolor{tableheadcolor}}c}{\boldmath$\order{(\alphas^2\alpha^2)^2}$}%
                & \multicolumn{1}{>{\columncolor{tableheadcolor}}c}{\boldmath$\order{(\alphas^3\alpha)^2}$}%
                & \multicolumn{1}{>{\columncolor{tableheadcolor}}c}{\bf Sum}%
                & \multicolumn{1}{>{\columncolor{tableheadcolor}}c}{\bf Total}\\ 
                \midrule
                $\Pg q$          & \multicolumn{1}{l}{--} &   0.231(4)  &   0.370(2)  & 0.365(1)   &   0.966 (4)  &   0.944 (9) \\
                $\Pg \bar{q}$    & \multicolumn{1}{>{\columncolor{tablerowcolor}}l}{--} &   0.0421(6) &   0.0679(3) & 0.0608(2)  &   0.1708(7)  &   0.167 (1) \\
                $q q^{(\prime)}$ &  0.001471(2)                                         &   0.0575(5) &   0.1106(2) & 0.07871(9) &   0.2483(6)  &   0.2478(8) \\
                $q \bar{q}$      &  0.01973(3)                                          &   2.531(6)  &   0.957(1)  & 0.00333(1) &   3.511 (6)  &   3.538 (4) \\
                $\Pg \Pg  $      & \multicolumn{1}{l}{--} &   8.01(2)   &  17.19(6)   & 0.00756(2) &  25.21  (6)   &  23.71 (6) \\\midrule\rowcolor{tableheadcolor}                                         
                $\sum$           &  0.02120(3)                                          &  10.87(2)   &  18.69(6)   & 0.516(2)   &  30.10  (6)   &  28.60 (6) \\
                \bottomrule
        \end{tabular}
        \caption{\label{table:results_full_matrixelement_summary} 
                Composition of the cross section in fb for
                the full process at the LHC at $13\TeV$. 
                Here $q q^{(\prime)}$ denotes pairs of quarks and/or
                antiquarks other than $q\bar{q}$.
        }
\end{table}
The results for the full process are listed in
\refta{table:results_full_matrixelement_summary}.  Here, additional
partonic channels ($\Pg q$, $\Pg \bar{q}$, $q q^{(\prime)}$) contribute about $5\,\%$.  The cross
section increases by merely $8\,\%$ relative to  $\ttbarbbbar$
production. The contributions of order  $\order{(\alphas^3\alpha)^2}$
are below $2\%$ and the interference pattern is similar to the
case of $\ttbarbbbar$
production.

In \citere{Denner:2014wka} we investigated the irreducible background
and interference effects for various distributions.  We found that
assigning two b jets to the top- and antitop-quark decay by maximising
a combined Breit--Wigner likelihood function and assigning the
remaining two b jets to the potential Higgs boson yields a good
unbiased determination of the b-jet pair originating from the
Higgs-boson decay.  While interference effects lead to a constant shift
in most of the differential distributions, they cause 
non-uniform shape changes in a few distributions like the one in the
invariant mass of the pair of bottom quarks not resulting from the top
quarks and the azimuthal angle between these bottom quarks.

\section{NLO QCD corrections to  $\Pp\Pp\to\Pe^+\nu_\Pe
\mu^-\bar{\nu}_\mu\Pb\bar{\Pb}\PH$}
\label{sec:NLO_results}

We have computed the NLO QCD corrections to the full hadronic process
\fullProcessH including all resonant, non-resonant, and off-shell
effects of the top quarks and all interferences at 
$13\TeV$ (for details see
\citere{Denner:2015yca}).  We include the tree-level amplitudes at
$\order{\alphas\alpha^{5/2}}$ for gluon-induced and
quark--antiquark-induced processes and the corresponding NLO
corrections of order $\alphas$. The bottom quark is considered
massless in this study.  The corresponding real corrections receive
also contributions of quark--gluon- and antiquark--gluon-initiated
processes.  We apply the Catani--Seymour subtraction formalism
\cite{Catani:1996vz,Catani:2002hc} for the regularization and
analytical cancellation of IR singularities.  We employ
\recola~\cite{Actis:2012qn} for the computation of all matrix elements
as well as colour- and spin-correlated squared matrix elements needed
for the evaluation of subtraction terms.

The matrix elements for the virtual corrections are calculated with
\recola in dimensional regularisation, which
uses the \collier~\cite{Denner:2014gla} library for the
numerical evaluation of one-loop scalar
\cite{'tHooft:1978xw,Beenakker:1988jr,Dittmaier:2003bc,Denner:2010tr}
and tensor integrals
\cite{Passarino:1978jh,Denner:2002ii,Denner:2005nn}.  We sucessfully
compared our results for the virtual NLO contribution to the squared
amplitude, $2\Re\mathcal{M}^*_0\mathcal{M}_1$, 
with \madgraph~\cite{Alwall:2014hca}. In addition we checked
the Ward identity for the matrix elements of the gluon-initiated
process at tree and one-loop level.

We use the anti-$k_\text{T}$ algorithm \cite{Cacciari:2008gp} for the
jet reconstruction with a jet-resolution parameter $R=0.4$. Only
final-state quarks and gluons with rapidity $|y|<5$ are clustered into
infrared-safe jets.
As default, we use a dynamical scale following \citere{Frederix:2011zi},
\begin{equation}\label{eqn:DynamicalScale}
        \mu_\text{dyn} = \mu_\text{R} = \mu_\text{F} = \left(m_{\text{T},\Pt}\,m_{\text{T},\bar{\Pt}}\, m_{\text{T},\PH}\right)^\frac{1}{3}\quad\text{with}\quad m_{\text{T}}=\sqrt{m^2+p^2_\text{T}}.
\end{equation}
Alternatively, we choose a fixed scale according to
\citere{Beenakker:2002nc} as given in \eqref{eqn:FixedScale}.
Scale uncertainties are determined by computing integrated and
differential cross sections at seven scale pairs,
$(\mu_\text{R}/\mu_0$,
$\mu_\text{F}/\mu_0)=(0.5,0.5),(0.5,1),(1,0.5),(1,1),(1,2),(2,1),(2,2)$.
The central value corresponds to $(\mu_\text{R}/\mu_0$,
$\mu_\text{F}/\mu_0)=(1,1)$, and the error band is constructed from the
envelope of these seven calculations.

\begin{table}
        \centering
        \renewcommand\arraystretch{1.2}
        \rowcolors{2}{tablerowcolor}{}
        \begin{tabular}{llll}
                \toprule\rowcolor{tableheadcolor}       
                  \multicolumn{1}{>{\columncolor{tableheadcolor}}l}{\textbf{pp}} 
                & \multicolumn{1}{>{\columncolor{tableheadcolor}}c}{\textbf{\boldmath$\sigma_\textbf{LO}$ [fb]}}%
                & \multicolumn{1}{>{\columncolor{tableheadcolor}}c}{\textbf{\boldmath$\sigma_\textbf{NLO}$ [fb]}}%
                & \multicolumn{1}{>{\columncolor{tableheadcolor}}c}{\boldmath$K$}\\
                \midrule            
                $\Pg \Pg$        &  $1.5906(1) ^{+33.7\%}_{-23.6\%}$ & $2.024(3) ^{+8.4\%  }_{-16.2\%}$   & 1.273(2)\\
                $q \bar{q}$      &  $0.67498(9)^{+24.1\%}_{-18.1\%}$ & $0.495(1) ^{+17.2\%  }_{-39.5\%}$  & 0.733(2)\\
                $\Pg \brabar{q}$ &                                   & $0.136(1) ^{+295\%   }_{-166\%}$ &     \\\midrule\rowcolor{tableheadcolor}
                $\sum$         &  $2.2656(1) ^{+30.8\%}_{-22.0\%}$ & $2.656(3) ^{+0.9\%   }_{-4.6\%}$   & 1.172(1)\\

                \bottomrule
        \end{tabular}
        \caption[Composition of the integrated cross section]{\label{table:results_summary}
                Composition of the integrated cross section for $\Pp\Pp \to 
                \Pe^+\nu_\Pe \mu^- \bar{\nu}_\mu \Pb \bar{\Pb}
                \PH(\Pj)$ 
                at the $13\TeV$ LHC
                with the dynamical  scale. In column one we list the partonic 
                initial states, where $q=\Pu,\Pd,\Pc,\Ps$ and $\brabar{q}=q,\bar{q}$. The second 
                and third column give the integrated cross sections in fb for LO and 
                NLO, resp., including scale uncertainties. The last column provides the $K$ 
                factor with $K=\sigma_\text{NLO}/\sigma_\text{LO}$.
        }
\end{table}
In \refta{table:results_summary} we present the integrated cross
sections for the dynamical scale \eqref{eqn:DynamicalScale}.
The cross sections for the fixed scale
\eqref{eqn:FixedScale} are lower by only about 1\,\%, and the
$K$~factor for the fixed scale is of 1.176(1). 
The contribution of the dominating gluon-fusion channel increases from
about 70\,\% at LO to 76\,\% at NLO.  The gluon--(anti)quark
induced real-radiation subprocesses contribute about 5\,\%. The
inclusion of NLO QCD corrections reduces the scale dependence from
\looseness -1
31\,\% to 5\,\%.
\begin{figure}
         \begin{center}
                \includegraphics[width=.485\linewidth]{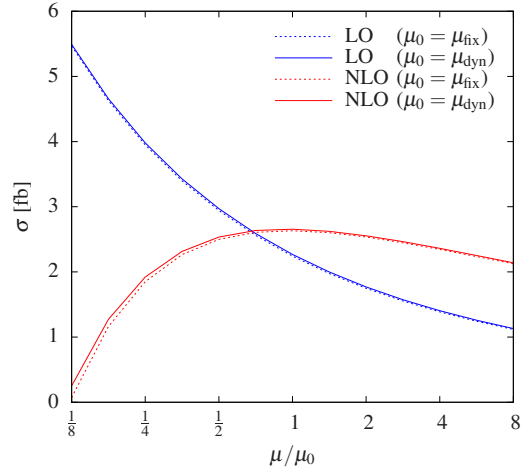}
                \caption{\label{plot:scale_dependence}%
                  Scale dependence of the LO and NLO integrated cross
                  section at the $13\TeV$ LHC.}
         \end{center}
\end{figure}
In \reffi{plot:scale_dependence} we display the dependence of the
integrated LO (blue) and NLO (red) cross sections on the values of the
fixed (dashed line) and dynamical scale (solid line) while keeping
$\mu_\text{R} = \mu_\text{F}$. Varying only $\mu_\text{R}$ or $\mu_\text{F}$
results in smaller variations.

The effects of the finite top-quark width have been determined via a
numerical extrapolation to the zero-top-width limit, $\Gamma_\Pt \to
0$.  For fixed scale $\mu_\text{fix}$
finite-top-width effects shift the LO and NLO cross section by
$-0.07\pm0.01\,\%$ and $-0.14\pm0.22\,\%$, respectively, which are
within the expected order of $\Gamma_\Pt/m_\Pt$. 

\begin{figure}
        \setlength{\parskip}{-10pt}
        \begin{subfigure}{0.50\textwidth}
                \subcaption{}
                \includegraphics[width=\textwidth]{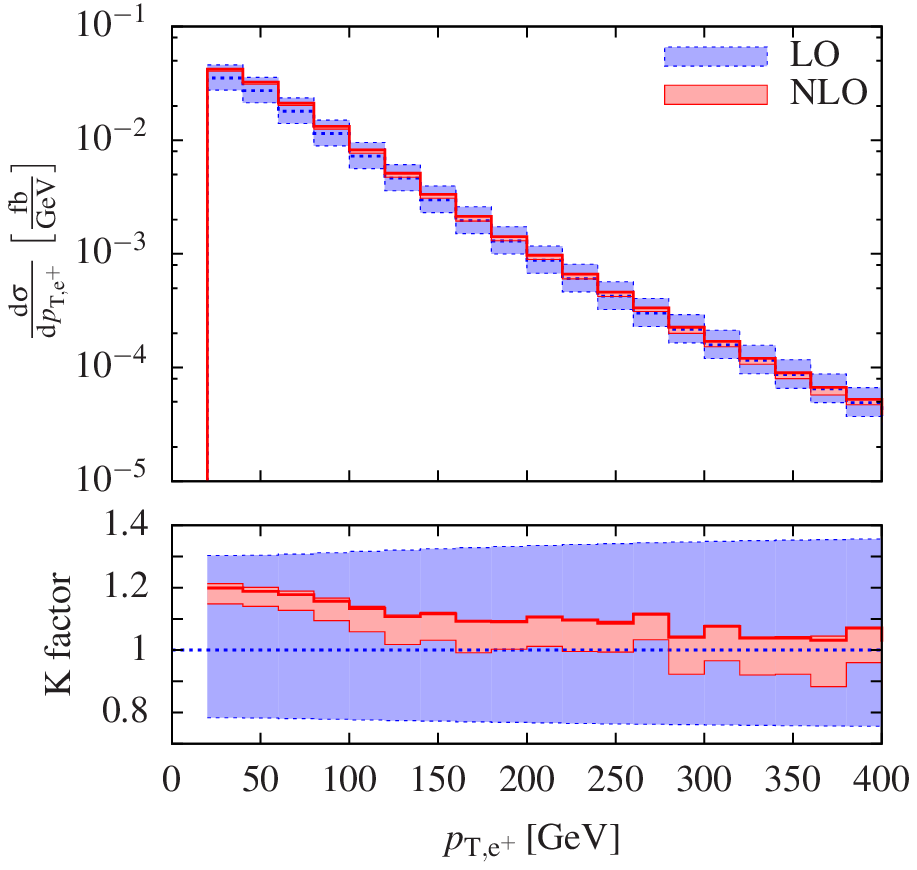}
                \label{plot:transverse_momentum_positron_dyn}
        \end{subfigure}
        \hfill
        \begin{subfigure}{0.5\textwidth}
                \subcaption{}
                \includegraphics[width=\textwidth]{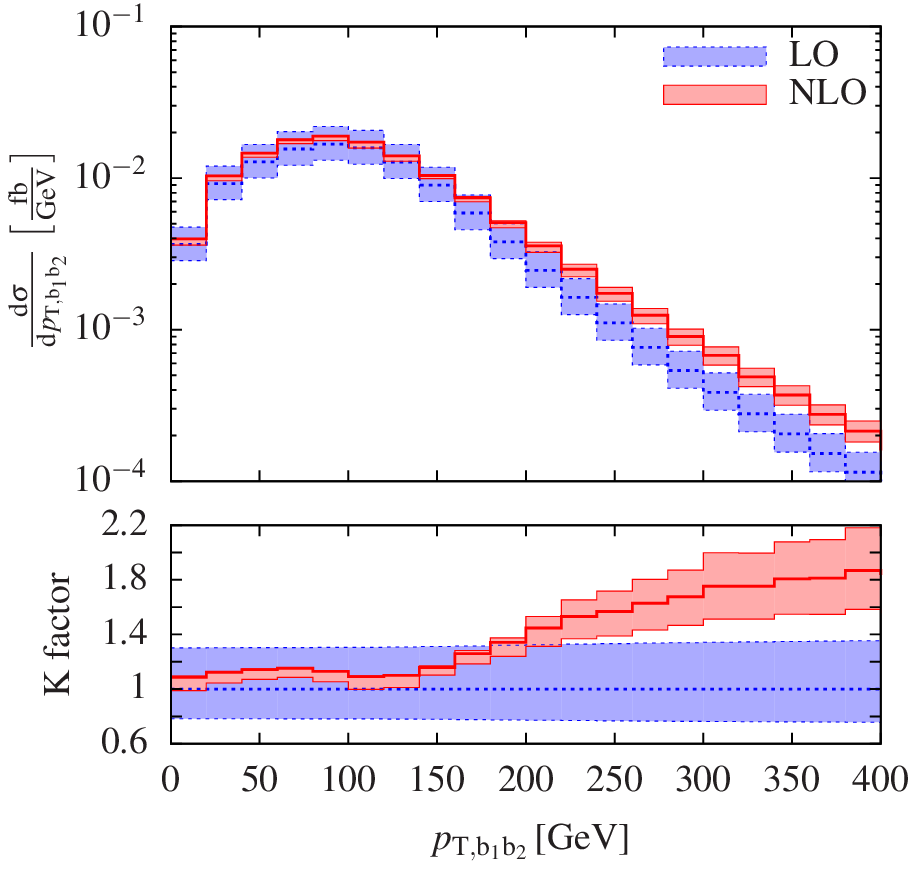}
                \label{plot:transverse_momentum_bb12_dyn}
        \end{subfigure}
        \vspace*{-3ex}
        \caption{\label{fig:transverse_momentum_distributions}%
                Transverse-momentum distributions at the
                $13\TeV$ LHC 
                 for the dynamical scale \protect\eqref{eqn:DynamicalScale}: 
                \subref{plot:transverse_momentum_positron_dyn} for the positron~(left), %
                \subref{plot:transverse_momentum_bb12_dyn} for the \Pb-jet pair~(right) and %
                The lower panels show the $K$~factor.}
\end{figure}%

We have investigated various differential distributions, two of which
are shown in \reffi{fig:transverse_momentum_distributions} for the
dynamical scale choice \eqref{eqn:DynamicalScale}.  The upper panels
show the LO (blue, dashed) and NLO (red, solid) predictions with
uncertainty bands from scale variations. The lower panels display the
LO (blue) and NLO (red) predictions with scale uncertainties
normalized to the LO results at the central scale.
Figure~\ref{plot:transverse_momentum_positron_dyn} shows the
transverse-momentum distribution of the positron. Using the dynamical
scale, the $K$~factor changes only slightly (within 20\,\%) over the
displayed range, and the NLO band lies within the LO band.  The
residual scale variation is at the level of $10\,\%$ at NLO.  This
behaviour is typical for most other distributions (see
\citere{Denner:2015yca}).
A notable exception is the distribution in the transverse momentum of
the \Pb-jet pair (\reffi{plot:transverse_momentum_bb12_dyn}), where we
observe an increase of the $K$ factor for high transverse momentum to
a value of 1.8 at $\pt\simeq400\GeV$. This originates from the fact
that this region is suppressed for on-shell top quarks, an effect
known already from $\Pt\bar\Pt$ production, where it is even more pronounced
\cite{Denner:2012yc}.

\section{Conclusions}
\label{sec:Conclusions}

We have investigated the irreducible background to the production of a
Higgs boson decaying into bottom quarks in association with a
top--antitop-quark pair including its decay. While $\Pt\bar\Pt\PH$
production contributes roughly a fourth to the $\Pl^+\nu_\Pl
\Pj\Pj\Pb\bar{\Pb}\Pb\bar{\Pb}$ final state, the major contribution of
$92\%$ is furnished by $\ttbarbbbar$ production. Interference effects
lower the corresponding cross section by about $5\,\%$. We have
calculated the next-to-leading-order QCD corrections to off-shell
top--antitop-quark production in association with a Higgs boson with
leptonic decay of the top quarks.  Using a dynamical scale, we find
$K$~factors mostly in the range $1.0{-}1.4$ and residual scale uncertainties
at the level of $10\,\%$ for distributions.

\acknowledgments 
This work was supported by the Bundesministerium
f\"ur Bildung und Forschung (BMBF) under contract no. 05H12WWE.

\providecommand{\href}[2]{#2}
\addcontentsline{toc}{section}{References}
\bibliographystyle{JHEP}
\bibliography{ttxh.bib}



\end{document}

%% file: macros.tex
\let\text\textrm

\newcommand{\Pl}{\ell}
\newcommand{\fb}{{\ensuremath\unskip\,\text{fb}}\xspace}

\def\reffi#1{\mbox{Figure~\ref{#1}}}

\def\refta#1{\mbox{Table~\ref{#1}}}
\def\reftas#1{\mbox{Tables~\ref{#1}}}

\def\citere#1{\mbox{Ref.~\cite{#1}}}
\def\citeres#1{\mbox{Refs.~\cite{#1}}}

\def\be{\begin{equation}}
\def\ee{\end{equation}}

\newcommand{\qqb}{\ensuremath{q\bar{q}}\xspace}
\newcommand{\PH}{\ensuremath{\text{H}}\xspace}
\newcommand{\Pj}{\ensuremath{\text{j}}\xspace}
\newcommand{\Pp}{\ensuremath{\text{p}}\xspace}
\newcommand{\Pe}{\ensuremath{\text{e}}\xspace}
\newcommand{\Pb}{\ensuremath{\text{b}}\xspace}

\newcommand{\Pt}{\ensuremath{\text{t}}\xspace}
\newcommand{\Pu}{\ensuremath{\text{u}}\xspace}
\newcommand{\Pd}{\ensuremath{\text{d}}\xspace}
\newcommand{\Ps}{\ensuremath{\text{s}}\xspace}
\newcommand{\Pc}{\ensuremath{\text{c}}\xspace}
\newcommand{\Pg}{\ensuremath{\text{g}}\xspace}

\newcommand{\PW}{\ensuremath{\text{W}}\xspace}
\newcommand{\PZ}{\ensuremath{\text{Z}}\xspace}

\newcommand{\Gt}{\ensuremath{\Gamma_\Pt}\xspace}

\newcommand{\GeV}{\ensuremath{\,\text{GeV}}\xspace}
\newcommand{\TeV}{\ensuremath{\,\text{TeV}}\xspace}

\newcommand{\alphas}{\ensuremath{\alpha_\text{s}}\xspace}
\newcommand{\order}[1]{\ensuremath{\mathcal{O}{\left(#1\right)}}\xspace}

\newcommand{\pt}{\ensuremath{p_\text{T}}\xspace}

\renewcommand{\Re}{\mathop{\mathrm{Re}}\nolimits}

\newcommand{\fullProcessH  }{\ensuremath{\Pp\Pp\to\Pe^+\nu_\Pe \mu^-\bar{\nu}_\mu\Pb\bar{\Pb}\PH}\xspace}

\newcommand{\recola}{{\sc Recola}\xspace}
\newcommand{\collier}{{\sc Collier}\xspace}
\newcommand{\madgraph}{{\sc\small MadGraph5\_aMC@NLO}\xspace}

\newcolumntype{.}{D{.}{.}{-1}}
\newcolumntype{d}[1]{D{.}{.}{#1}}

\colorlet{tableoverheadcolor}{gray!37.5}
\colorlet{tableheadcolor}{gray!25}
\colorlet{tablerowcolor}{gray!12.5}

\newlength{\width}
\newlength{\height}
\newcommand{\brabar}[1]{%
    \settoheight{\height}{\ensuremath{#1}}%
    \settowidth{\width}{\ensuremath{#1}}%
    \makebox[0pt][l]{\ensuremath{#1}}%
    \raisebox{1.26ex}{\scalebox{.3}{\textbf{(}}}%
    \rule[1.41\height]{0.7\width}{0.35pt}%
    \raisebox{1.26ex}{\scalebox{.3}{\textbf{)}}}%
}

\newcommand{\ttbarbbbar}{\ensuremath{\Pt\bar{\Pt}\Pb\bar{\Pb}}\xspace}
\newcommand{\ttbarh}{\ensuremath{\Pt\bar{\Pt}\PH}\xspace}
\newcommand{\ttbarhProcess}{\ensuremath{\Pp\Pp\to\Pt\bar{\Pt}\PH\to\Pl^+\nu_\Pl \Pj\Pj\Pb\bar{\Pb}\Pb\bar{\Pb}}\xspace}
\newcommand{\ttbarProcess }{\ensuremath{\Pp\Pp\to\Pt\bar{\Pt}\Pb\bar{\Pb}\to\Pl^+\nu_\Pl \Pj\Pj\Pb\bar{\Pb}\Pb\bar{\Pb}}\xspace}
\newcommand{\fullProcess  }{\ensuremath{\Pp\Pp\to\Pl^+\nu_\Pl \Pj\Pj\Pb\bar{\Pb}\Pb\bar{\Pb}}\xspace}